\documentclass{Odyssey2026}

\usepackage{stfloats}
\usepackage[table]{xcolor}

\usepackage{graphicx}
\usepackage{booktabs}

\usepackage{url}

\usepackage{multirow}
\usepackage{caption}
\usepackage{subcaption}
\usepackage{makecell}

\usepackage{setspace} 

\interspeechcameraready 

\title{Spoken Language Identification with Pre-trained Models and Margin Loss\thanks{$^*$Corresponding author. $^\dagger$ Project leader.}}

\author[affiliation={1,4}]{Zhihua}{Fang}
\author[affiliation={1,2,3,*}]{Liang}{He}
\author[affiliation={4,\dagger}]{Weiwu}{Jiang}

\affiliation{School of Computer Science and Technology}{Xinjiang University}{Urumqi, China}
\affiliation{School of Intelligence Science and Technology}{Xinjiang University}{Urumqi, China}
\affiliation{Department of Electronic Engineering}{Tsinghua University}{Beijing, China}
\affiliation{Central R\&D Department}{AGIBOT}{Shanghai, China}

\email{fangzhihua@stu.xju.edu.cn, heliang@mail.tsinghua.edu.cn, jiangweiwu@agibot.com}


\usepackage{comment}
\usepackage{soul}
\begin{document}

\maketitle

\begin{abstract}
For the speaker-controlled spoken language identification task proposed in the TidyLang Challenge 2026, this paper proposes a language identification method based on pre-trained models and margin-based losses. The proposed method adopts a pre-trained ECAPA-TDNN as the feature encoder and incorporates margin-based losses to enhance the discriminative ability of language representations, thereby improving inter-class separability and reducing the interference of non-linguistic factors such as speaker characteristics.
Experimental results on the Tidy-X dataset show that the proposed method achieves 85.95\% macro accuracy and 90.96\% micro accuracy on the language identification task and 17.08\% equal error rate (EER) on the verification task. Compared with the official baseline, the macro accuracy improves by 45.7\%, the micro accuracy improves by 15.2\%, and the EER is reduced by approximately 50.8\%, demonstrating the effectiveness of the proposed method. The code will be released at 
GitHub\footnote{\href{https://github.com/PunkMale/TidyLang2026}
{\mbox{\textcolor[rgb]{0.874, 0.0, 0.486}{\nolinkurl{https://github.com/PunkMale/TidyLang2026}}}}}.
\end{abstract}

\section{Introduction}
Spoken language identification (SLID) aims to automatically determine the language of an input speech signal, and is a fundamental task in audio signal processing, with important applications in automatic speech recognition front-ends, multilingual speech interaction, and multilingual speech retrieval~\cite{slid_overview2025speechcommun}. Traditional language identification tasks usually assume that speaker identity is only a nuisance factor, and the focus is on extracting stable language cues from acoustic and prosodic features. However, in real multilingual environments, the same speaker often switches between multiple languages across different contexts, which makes models more likely to rely on speaker-specific characteristics rather than language itself, leading to the problem of ``shortcut learning''~\cite{shortcut_learning2020nmi}. Based on this background, the TidyLang Challenge 2026 proposes a speaker-controlled spoken language identification task~\cite{tidylang2026}, whose core difference lies in explicitly emphasizing the disentanglement of speaker and language information, and further evaluating the generalization ability of models to unseen languages.

The TidyLang challenge 2026 includes two related tasks. The first is a language identification task on seen languages, where the system is required to perform language classification within the training language set, and macro accuracy is used as the main evaluation metric. The second is an unseen language verification task, where the system is required to make decisions based only on the similarity between enrollment speech and test speech, and equal error rate is used as the main metric. Compared with traditional closed-set language classification, this setting is closer to real-world applications and poses higher requirements on language representation ability, speaker robustness, and cross-lingual generalization. The public baseline results indicate that this task remains challenging, especially in terms of macro-level performance and unseen language recognition.

Research on spoken language identification has evolved from statistical modeling to deep representation learning. Early methods mainly relied on acoustic features combined with modeling frameworks such as Gaussian Mixture Model (GMM)~\cite{spk_gmm1995speechcommun}, Support Vector Machine (SVM)~\cite{spk_svm2006spl}, or i-vector~\cite{lid_i_vector2011interspeech}. Later, end-to-end methods based on deep neural networks gradually became dominant, where x-vector~\cite{slid_x_vector2018odyssey}, Time-Delay Neural Network (TDNN)~\cite{slid_tdnn2016interspeech}, and their variants achieved good performance in both language and speaker recognition tasks. Among them, ECAPA-TDNN~\cite{ecapatdnn2020interspeech} improves discriminative representation ability through multi-scale feature aggregation and channel attention mechanisms, showing strong robustness in related tasks. In addition, self-supervised pre-trained models such as wav2vec 2.0~\cite{lid_wav2vec2021interspeech, lid_wav2vec2023interspeech}, HuBERT~\cite{hubert2021taslp}, and XLS-R~\cite{xls_r2022interspeech} provide new approaches for low-resource and cross-lingual speech modeling. However, under the speaker-controlled language identification setting, how to leverage pre-trained knowledge while suppressing speaker-related interference remains an open problem.

To address the above issues, this paper proposes a spoken language identification method based on pre-trained models and margin-based losses. Specifically, we adopt an ECAPA-TDNN model pre-trained on VoxLingua107 as the feature encoder, and introduce margin-based loss functions to enhance the discriminative ability of language representations, thereby improving inter-class separability and suppressing the interference of non-linguistic factors such as speaker characteristics. In addition, we conduct a systematic comparison between Additive Angular Margin Softmax~\cite{arcface2019cvpr} and Real Additive Margin Softmax~\cite{ram_softmax2022icassp}, and introduce a self-supervised pre-trained model, XLS-R~\cite{xls_r2022interspeech}, as an alternative encoder to analyze the impact of different pre-training strategies on system performance. In summary, the main contributions of this paper are as follows:
\begin{enumerate}
    \item We propose a spoken language identification framework based on pre-trained models and margin-based losses, which significantly outperforms the official baseline.
    \item We compare ECAPA-TDNN and XLS-R as encoders, and verify the advantage of task-related pre-training for the SLID task.
    \item We analyze the performance differences between AAM-Softmax and RAM-Softmax in both classification and verification tasks, providing empirical insights into the application of margin-based losses for language identification.
\end{enumerate}

The remainder of this paper is organized as follows. Section~\ref{sec:preliminaries} introduces the TidyLang Challenge 2026 and its baseline system. Section~\ref{sec:method} describes the proposed method in detail. Section~\ref{sec:experiment} presents the experimental setup and results analysis. Section~\ref{sec:conclusion} concludes the paper and discusses future work.

\section{Preliminaries}
\label{sec:preliminaries}
\subsection{Challenge Description and Dataset}
The TidyLang Challenge 2026 focuses on the problem of speaker-controlled spoken language identification. Unlike traditional language identification tasks that usually treat speaker identity as an interfering factor, this challenge explicitly focuses on the scenario where “the same speaker uses multiple languages,” requiring the system to perform discrimination as much as possible based on language-related cues rather than speaker characteristics, and further evaluates the model’s generalization ability to unseen languages. The challenge is evaluated based on the Tidy-X dataset~\cite{tidyvoice2026arxiv}, which is a multilingual corpus organized from Mozilla Common Voice~\cite{common_voice2020lrec}, containing more than 4,474 speakers, 40 languages, about 321,711 utterances, and approximately 457 hours of speech data, with each speaker corresponding to speech samples in 2–10 languages.

The challenge consists of two related tasks. Task 1 is seen-language identification, which performs language classification for each utterance across 35 training languages, with evaluation metrics of macro- and micro-accuracy. Task 2 is unseen-language identification, which targets 40 languages not seen during training and adopts a verification approach based on enrollment speech. The system is required to output the similarity or probability score between enrollment and test speech, with the main evaluation metric being Equal Error Rate (EER). In addition, the challenge defines both closed-condition and open-condition settings: in the closed-condition, only the officially provided Tidy-X training and validation data are allowed, whereas in the open-condition, other language identification data resources may be used without requiring additional Common Voice data.

\subsection{Baseline System}
The organizers provide a baseline system\footnote{\url{https://github.com/areffarhadi/TidyLang2026-baseline}} along with supporting training and evaluation scripts. This framework uses Wav2Vec2-Large~\cite{wav2vec2020neurips} as the acoustic backbone, extracts representations from layers 17–24 and aggregates them through learnable weights; then a lightweight projection head is used to map the features into 256-dimensional embeddings with L2 normalization; in the classification stage, the ArcFace loss~\cite{arcface2019cvpr} is adopted for training, with default parameters of margin 0.3 and scale 30.0. The input audio sampling rate is 16 kHz, and each utterance is about 4 seconds long.

\section{Method}
\label{sec:method}
\subsection{Pre-trained ECAPA-TDNN Encoder}
We adopt a pre-trained ECAPA-TDNN~\cite{ecapatdnn2020interspeech} as the speech encoder for spoken language identification. Built upon the TDNN architecture, ECAPA-TDNN introduces stronger channel modeling, multi-scale temporal modeling, and attentive statistics pooling, and therefore can learn more robust utterance-level representations~\cite{lid_ecapatdnn2023aic}.

In recent years, pre-training on large-scale speech data followed by fine-tuning on downstream tasks has become a common practice in speech signal processing tasks. Compared with training from scratch, pre-trained models usually provide better parameter initialization, faster convergence, and more stable representation learning. In spoken language identification, this pre-training–fine-tuning paradigm has also been shown to be effective and can significantly improve the modeling ability of language features~\cite{lid_wav2vec2023interspeech}.

For the speaker-controlled spoken language identification task, this pre-training–fine-tuning strategy is particularly important. On the one hand, it helps the model learn language-related acoustic patterns; on the other hand, it reduces overfitting to non-linguistic factors such as speaker identity and recording conditions. Therefore, we adopt an ECAPA-TDNN\footnote{\url{https://huggingface.co/speechbrain/lang-id-voxlingua107-ecapa}} model pre-trained on VoxLingua107~\cite{voxlingua2021slt} as the initial encoder and fine-tune it on the Tidy-X\footnote{\url{https://datacollective.mozillafoundation.org/datasets/cmihtsewu023so207xot1iqqw}} training set.

\subsection{Margin-based Classification Heads}
Cross-entropy (CE) loss is the most commonly used objective function for multi-class classification, but it mainly focuses on classification correctness and does not explicitly enforce the margin between the target class and non-target classes. In spoken language identification, using CE loss alone may lead to learned representations that lack sufficient discriminative power, especially when speaker interference is strong or language boundaries are ambiguous.

To enhance the discriminative ability of language representations, margin-based softmax losses have been widely adopted in recent years. The core idea is to introduce an explicit margin between the target class and its neighboring classes, thereby pushing the decision boundary farther away. In this work, we explore the effectiveness of Additive Angular Margin Softmax~\cite{arcface2019cvpr} (AAM-Softmax) and Real Additive Margin Softmax~\cite{ram_softmax2022icassp} (RAM-Softmax) for the spoken language identification task, which are described in detail below.

\subsubsection{Additive Angular Margin Softmax}
AAM-Softmax introduces an additive angular margin in the angular space. Its core idea is to impose an additional angular constraint on the target class, thereby enlarging the discriminative margin between the target class and non-target classes. Compared with the standard softmax, it replaces $\cos\theta_{i,y_i}$ with $\cos(\theta_{i,y_i}+m)$ for the target class, requiring stricter angular discrimination during classification. Its formal definition is as follows:
\begin{equation}
L_\mathrm{AAM} = -\frac{1}{N} \sum_{i=1}^{N}
\log \frac{
e^{s \cos(\theta_{y_i} + m)}
}{
e^{s \cos(\theta_{y_i} + m)} +
\sum_{j=1, j \neq y_i}^{n}
e^{s \cos \theta_j}
},
\end{equation}
where $N$ denotes the number of samples in a mini-batch, $C$ denotes the number of classes, $y_i$ is the ground-truth label of the $i$-th sample, $\theta_{i,j}$ represents the angle between the feature vector of the $i$-th sample and the weight vector of class $j$, $s$ is the scale factor, and $m$ is the additive angular margin.

By explicitly introducing an angular margin on the target class, this loss encourages more compact intra-class distributions and better separated inter-class distributions. Therefore, it has been widely used in speaker recognition and language recognition tasks. For the task considered in this work, AAM-Softmax helps reduce the interference caused by speaker variation and improves the separability among language classes.

\subsubsection{Real Additive Margin Softmax}
RAM-Softmax revisits the definition of margin from the perspective of a “real margin”. Its basic idea is that the discriminative ability of the model should be measured by the actual difference between the target logit and all non-target logits, rather than by simply imposing an intuitive margin on the target class. Based on this idea, RAM-Softmax applies additional penalties only to hard non-target classes, while easy non-target classes that have already been sufficiently separated are no longer further optimized. It is defined as:
\begin{equation}
L_{\mathrm{RAM}}
=
\frac{1}{N}\sum_{i=1}^{N}
\log\left\{
1+
\sum_{\substack{j=1\\j\neq y_i}}^{C}
e^{\max\left\{0,\,-s\left(\cos\theta_{i,y_i}-\cos\theta_{i,j}-m\right)\right\}}\right\}.
\end{equation}
The hyperparameter definitions are almost the same as those in AAM-Softmax.
RAM-Softmax directly models the real margin between the target class and non-target classes. When the target class already satisfies a sufficient margin over a certain non-target class, the contribution of that non-target class to the loss is suppressed; when a non-target class is still close to the target class, its corresponding term is amplified, thereby imposing stronger optimization pressure on the model. In spoken language identification (SLID), this mechanism helps the model focus on fine-grained differences between similar languages, thus improving inter-class separability. Therefore, RAM-Softmax emphasizes hard negatives and helps to learn more discriminative speech embeddings.

\subsection{Overall Framework}
Our overall framework consists of a pre-trained ECAPA-TDNN encoder and a margin-based classification head. The input speech is first processed by the pre-trained ECAPA-TDNN to extract utterance-level representations, which are then fed into an AAM-Softmax or RAM-Softmax head for loss computation and training.

For the language identification task, we directly use the class scores produced by the classification head for prediction, and select the class with the highest score as the final result. For the open-set language verification task, we only use the encoder to extract features from enrollment and test utterances, and compute scores using cosine similarity.

\section{Experiments}
\label{sec:experiment}
\begin{table*}[t]
    \centering
    \caption{Main results on Task 1 and Task 2 of the TidyLang Challenge 2026.}
    \label{tab:main_results}
    \begin{tabular}{l l l c c c}
    \hline
    System & Encoder & Loss Function & Macro Accuracy (\%) $\uparrow$ & Micro Accuracy (\%) $\uparrow$ & EER (\%) $\downarrow$\\
    \hline
    Baseline & Wav2Vec2-Large & AAM-Softmax & 40.25 & 75.76 & 34.70\\
    \hline
    \multirow{3}{*}{Ours}
    & XLS-R & AAM-Softmax & 65.71 & 81.63 & - \\
    & ECAPA-TDNN & AAM-Softmax & \textbf{85.95} & 90.96 & 17.08 \\
    & ECAPA-TDNN & RAM-Softmax & 85.91 & \textbf{91.73} & \textbf{16.39} \\
    \hline
    \end{tabular}
\end{table*}

\subsection{Experimental Details}
We only participate in the closed-condition track of the TidyLang Challenge 2026, where the model is trained using only the provided Tidy-X dataset. Under this condition, we report the results on both Task 1 and Task 2. For Task 1, macro accuracy and micro accuracy are used as evaluation metrics, while for Task 2, equal error rate (EER) is used. We strictly follow the official data split and evaluation protocol for training and testing.

We train the model on a single NVIDIA RTX 4090D GPU. The maximum number of training epochs is set to 30, and the batch size is 64. The initial learning rate is set to $1\times10^{-4}$ and is decayed using a cosine annealing schedule. The AdamW~\cite{adamw2019iclr} optimizer is used. The margin $m$ and scale $s$ for both AAM-Softmax and RAM-Softmax are set to 0.2 and 30, respectively. In addition to using an ECAPA-TDNN model pre-trained on VoxLingua107 as the encoder, we also experiment with XLS-R\footnote{\url{https://huggingface.co/facebook/wav2vec2-xls-r-300m}}, which is pre-trained on 436,000 hours of unlabeled speech data, as an alternative encoder.

We apply data augmentation to the training speech with a probability of 0.8. Specifically, the MUSAN~\cite{musan2015} dataset is used to simulate noise, music, and speech interference, while the RIRS~\cite{rirs2017icassp} dataset is used to simulate reverberation. By randomly introducing noise and reverberation, the model is exposed to more complex acoustic conditions, which improves its robustness to noise and environmental variations in real-world scenarios. In the SLID task, this augmentation strategy helps the model learn more robust language-related features, thereby reducing its reliance on speaker and environmental factors.

\subsection{Results and Analysis}
Table~\ref{tab:main_results} presents the comparison between our systems and the official baseline on Task 1 and Task 2. It can be seen that our methods significantly outperform the official baseline on both language identification and language verification tasks. For the language identification task, both macro accuracy and micro accuracy are greatly improved; for the language verification task, the EER is significantly reduced from 34.70\% to 17.08\% and 16.39\%, demonstrating that the proposed method also has clear advantages in open-set discrimination.

From the encoder comparison, XLS-R already brings a noticeable improvement over the official baseline, indicating that self-supervised pre-trained speech models are effective for the SLID task. However, compared with ECAPA-TDNN, XLS-R still shows a large gap in both macro accuracy and micro accuracy. This suggests that, compared with general self-supervised speech representations, encoders pre-trained for SLID-related tasks are more suitable for the current challenge and can learn more discriminative language-related information. Meanwhile, from the verification results, the ECAPA-TDNN-based systems achieve a significant reduction in EER, further validating their generalization ability in open-set scenarios.

From the loss function comparison, under the same ECAPA-TDNN encoder, RAM-Softmax achieves similar macro accuracy to AAM-Softmax, but performs better in both micro accuracy and EER. Specifically, RAM-Softmax improves Micro Accuracy from 90.96\% to 91.73\%, and reduces EER from 17.08\% to 16.39\%. This indicates that RAM-Softmax focuses more on hard negatives during optimization, which helps improve performance in fine-grained classification and open-set verification tasks.

Overall, the choice of pre-trained speech encoder has a more significant impact on performance. Compared with general self-supervised pre-trained speech models, the SLID-related pre-trained ECAPA-TDNN shows clear advantages in this challenge. On top of that, an appropriate margin-based loss function can further bring performance gains.

\section{Conclusion}
\label{sec:conclusion}
This paper investigates spoken language identification with pre-trained models and margin-based losses for the speaker-controlled spoken language identification task in the TidyLang Challenge 2026. The experimental results show that the ECAPA-TDNN pre-trained on VoxLingua107 significantly outperforms both the official baseline and the self-supervised pre-trained speech model. Moreover, RAM-Softmax further improves the overall performance compared with AAM-Softmax, demonstrating the effectiveness of real-margin-based loss design for this task.

Despite the promising results, several limitations remain. First, this work mainly focuses on Task 1, and the system design and optimization for unseen-language recognition are still insufficient. Second, the potential of self-supervised pre-trained models has not been fully explored and may be further improved. In addition, the current approach relies on a single encoder architecture without investigating model fusion or more advanced modeling strategies. Future work will extend to unseen-language recognition, explore stronger pre-training strategies and loss designs, and investigate more robust representation learning methods to further improve spoken language identification under speaker-controlled conditions.

\section{Acknowledgments}
This work was supported by the National Natural Science Foundation of China under Grant No. 62366051.

\bibliographystyle{IEEEtran}
\bibliography{reference}

\end{document}